\documentclass[12pt]{iopart}

\usepackage{graphicx}
\begin{document}

\title{Colossal Enhancement of Electrical Conductivity in $Y_2Ir_2O_7$ Nanoparticles}

\author{Vinod Kumar Dwivedi \textit{$^{1}$}, Abhishek Juyal \textit{$^{2}$}, \& Soumik Mukhopadhyay \textit{$^{2}$}}

\address{\textit{$^{1}$} Materials Science Programme, Indian Institute of Technology Kanpur, Kanpur 208016, India}

\address{\textit{$^{2}$} Physics Department, Indian Institute of Technology Kanpur, Kanpur 208016, India}
\ead{vinodd@iitk.ac.in}
\begin{abstract}

We present a comparative study of the magnetic and electrical properties of polycrystalline and nanocrystalline $Y_2Ir_2O_7$, the latter prepared using a new chemical route. We find that reduction in particle size leads to enhanced ferromagnetism and orders of magnitude enhancement of electrical conductivity in the nanocrystalline sample. Based on X-ray photoelectron spectroscopy and X-ray absorption near edge structure spectroscopy results, the phenomenon is attributed to the increased fraction of $Ir^{5+}$ and the resulting increase in double exchange interaction along with the weakening of spin-orbit coupling strength in the nanocrystalline sample compared to the bulk.
\end{abstract}


\vspace{2pc}

%
%

\section{Introduction}

In 4d and 5d transition metal oxides (TMO), an intermediate correlation effect appears due to the interplay between comparable energy scales, namely, the relativistic spin orbit coupling, on site coulomb repulsion and crystal field effect despite the extended nature of the one-electron wave functions. This provides us with an opportunity to manipulate the balance between these energy scales with small perturbation leading to dramatic effect arising from the competing spin, orbital, charge, and lattice degrees of freedom~\cite{Wan,Kim,Pesin,Clancy,Yang,Bhattacharjee,William}. Among all the 5d TMOs, Pyrochlore Iridates $R_2Ir_2O_7$ (R = Yttrium,or lanthanide elements) are proving to be the most interesting because of the possibility of realization of a variety of novel phases ~\cite{Kim,Pesin,Wan} such as topological Mott insulators ~\cite{Pesin,Yang,Kargarian}, chiral spin liquids~\cite{Machida}, Weyl semimetals~\cite{Wan,William}, and axion insulators~\cite{Wan,Go}, which can be accessed by tuning the relative strengths of the relevant energy scales.

In Pyrochlore Iridates, the electronic and magnetic properties are expected to be strongly coupled to each other ~\cite{Wan,Chen,William1}. The magnetic $Ir^{4+}$ ions are distributed over a network of corner sharing tetrahedra and each of the four Ir atoms is coordinated by six oxygen atoms accompanied by trigonal compression of the oxygen cage along [111] direction~\cite{William}. In general, the magnetic ground state emerges due to the interplay between Heisenberg-type antiferromagnetic interaction, Dzyaloshinskii-Moriya interaction emerging due to strong spin-orbit coupling and single-ion anisotropy~\cite{Wan,Shinaoka,Bogdanov}. If $R^{3+}$ is also magnetic, more complex magnetic ground state could be obtained~\cite{Guo,Disseler2,Tomiyasu,Taira,MacLaughlin}. Fortunately, the problem simplifies considerably for $Y_2Ir_2O_7$ since it has nonmagnetic $Y^{3+}$ ions residing at R-site.
 It was theoretically predicted that $Y_2Ir_2O_7$ could have a unique all-in/all-out antiferromagnetic (AFM) ground state~\cite{Wan} with the low -lying excited state being ferromagnetic. Experimentally, although there is a consensus regarding the existence of long range magnetic ordering~\cite{Disseler}, the precise nature of the ordering is not yet fully established, primarily due to the limitation of neutron diffraction studies in iridates~\cite{Shapiro}. Probabilistic modeling and ab initio calculations based on experimentally measured spontaneous muon spin precession frquency do, however, suggest that $Y_2Ir_2O_7$ indeed has all-in/all-out antiferromagnetic ground state~\cite{Disseler1}. Interestingly, there are some reports which point towards a ferromagnetic component on top of AFM ground state~\cite{Shapiro,Zhu} based on observation of hysteresis in M-H data at low temperature. So far as electronic property is concerned, although high resolution photoemission analysis reveals finite density of state at the Fermi level~\cite{Singh}, it is suggested that bulk $Y_2Ir_2O_7$ could be a Mott insulator~\cite{Taira,Fukazawa,Aito,Soda,Yanagishima}.

Surprisingly, till date, there is no report regarding the influence of particle size reduction on the properties of Pyrochlore Iridates. This is particularly important as regards the applicability of Mott physics in 5d oxide systems in general in that long-ranged interactions and cooperative phenomena should be significantly influenced by size effects. In this work, $Y_2Ir_2O_7$ nanoparticles were synthesized for the first time. We present an extensive comparative analysis of the magnetic and electrical properties of $Y_2Ir_2O_7$ in nanocrystalline and polycrystalline form using dc magnetization and electrical transport measurements along with X-ray photoelectron spectroscopy (XPS) and X-ray absorption near edge structure (XANES) spectroscopy measurements.

\section{Experimental methods}

Nanocrystalline $Y_2Ir_2O_7$ sample was prepared by chemical route. High purity stoichiometric solutions of yttrium oxide $Y_2O_3$ (SIGMA-ALDRICH, 99.99\%) and iridium acetate $IrC_6H_9O_6$ [Alfa-Aesar, Ir ($48 - 54\%$)] were used for synthesis. To dissolve $Y_2O_3$ 50\% concentrated $HNO_3$ was used whereas $IrC_6H_9O_6$ was dissolved in DI water.  The solution was heated at 50$^o$C for 2 hours in a magnetic stirrer. The resulting solution was heated at 80$^o$C for 12 hours and a black colored gel was obtained, which was then calcined at 600$^oC$ for 6 hours in air. The calcined powder was reground and pressed into pellet  and heated at 900$^o$C for 4 hours. The resulting pellet was crushed, reground and pelletized into 2 mm disk of diameter of 6 mm under applied pressure of 50 kPa/cm$^2$.  The final sintering was done at 1000$^o$C for 6 hours. On the other hand, pollycrystalline $Y_2Ir_2O_7$ was made by conventional solid state reaction route. Mixtures of $Y_2O_3$ (SIGMA-ALDRICH) and $IrO_2$ [Alfa-Aesar, Permion (r)] with purities of 99.99\% were used. The molar ratio of Y/Ir was fixed at 1:1.1. These mixtures  were ground, pelletized, and then heated in air at 1000$^o$C for 100 hours. The resulting material was reground, pressed into pellets of same dimension and applied pressure as used for nanocrystalline sample, and resintered at the same temperature for an additional 150 hours with two intermediate regrindings.

Structural properties and phase formation were studied using X-ray diffraction (XRD) using a PANalytical X’PertPRO diffractometer with CuK$\alpha$ radiation ($\lambda= 1.54056 \AA$). The particle size, distribution and crystallinity were verified using transmission electron microscopy (TEM). Particle size distribution was also verified using Field emission scanning electron microscope (FE-SEM) JSM-7100F, JEOL. The chemical composition of the sample was determined using energy dispersive X-ray spectrometry (EDX). The temperature dependence of zero-field-cooled and field-cooled magnetization with applied magnetic field of 1 kOe in the temperature range 2-300 K and Magnetic field dependent magnetization loop with maximum field 10 T at low temperature were measured using a Quantum Design MPMS SQUID magnetometer. Electrical resistivity and transverse magnetoresistance of the samples were measured by conventional four probe technique using a Quantum Design PPMS. X-ray photoelectron spectroscopy (XPS) was performed using a PHI 5000 Versa Probe II system to determine the electronic structure.

X-ray absorption near edge structure (XANES) measurements were carried out in the transmission mode at Ir $L_{3,2}$ edge for the nanocrystalline sample, taken in powder form and mixed with cellulose powder to obtain total weight of approximately 100 mg and 2.5 mm thick homogenous pellets of 15 mm diameter. XANES measurements were done at beamline (BL-9) at the INDUS-2 Synchrotron Source (2.5 GeV, 100 mA) at the RRCAT, Indore, India~\cite{http}. The beamline have photon energy ranging from 4–25 keV and resolution of 10000 at 10 keV.

\section{Results and Discussion}
\begin{figure}[h]
\centering
\includegraphics[width=10cm]{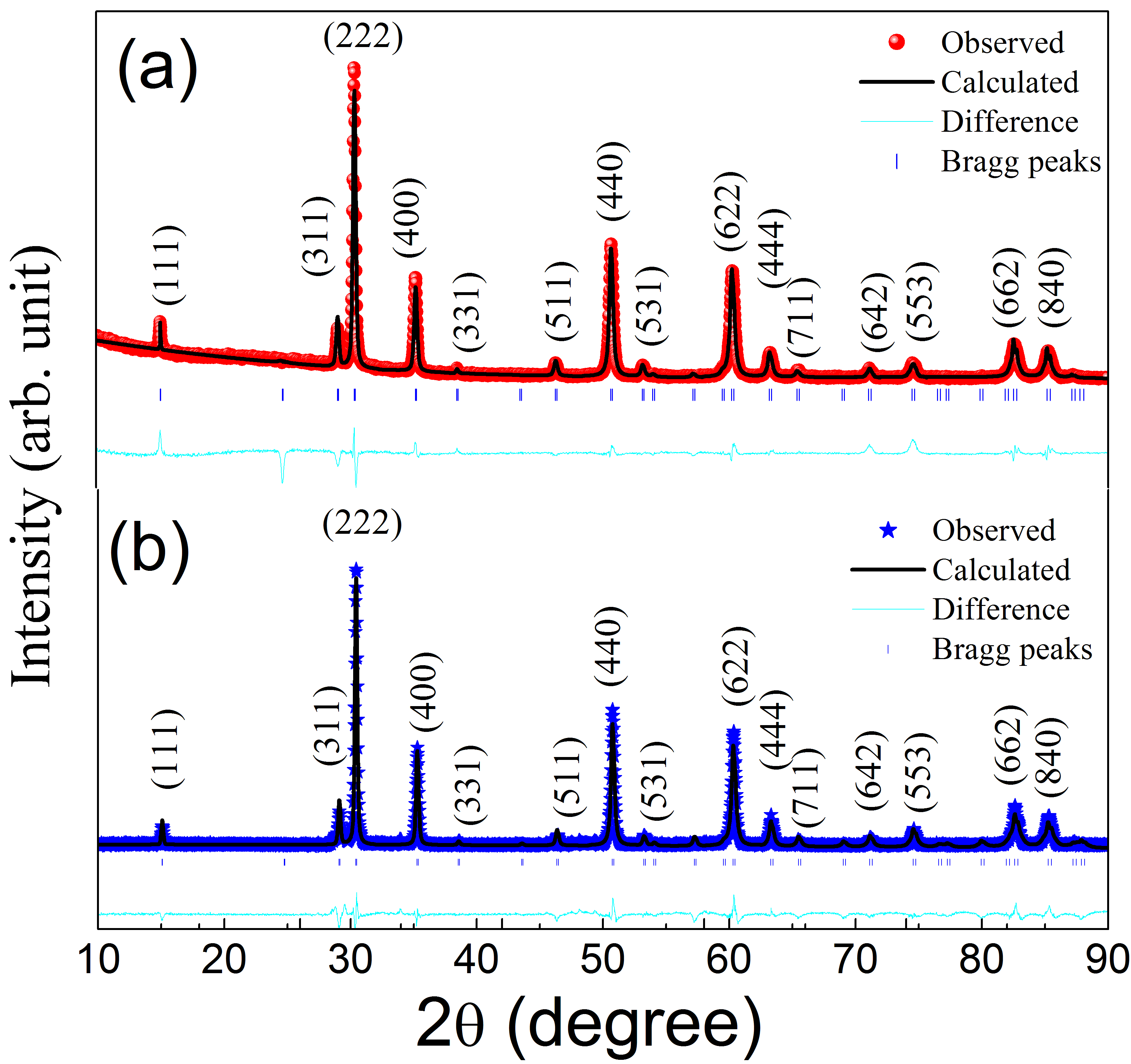}
\caption{Powder X-ray diffraction patterns of (a) nanocrystalline and (b) polycrystalline $Y_2Ir_2O_7$ measured at room temperature (RT). Observed and refined data are shown by red solid circle (nanocrystalline) and blue solid star (polycrystalline) and black solid curves, respectively. Vertical bars represent positions of the Bragg reflections. The solid line at the bottom shows deviation between the experimental results and the calculation.}\label{fig:xrd}
\end{figure}
\begin{figure}[h]
\centering
\includegraphics[width=12cm]{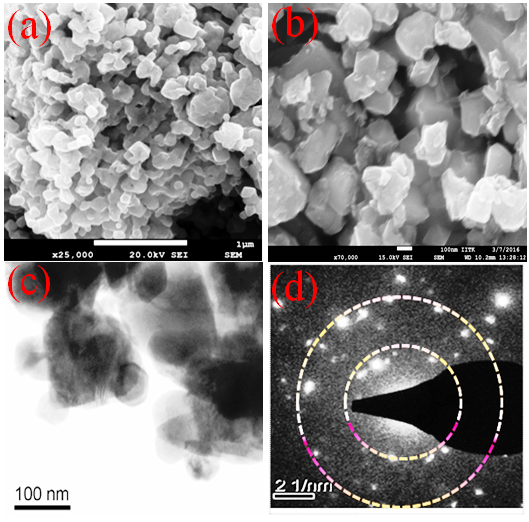}\\
\caption{Field emission Scanning electron microscope (FE-SEM) micrograph of nanocrystalline (a) and polycrystalline (b) samples. Transmission electron microscope (TEM) micrograph (c) show 50 nm  average particle size and Selected Area Electron Diffraction (SAED) pattern (d) of nanocrystalline $Y_2Ir_2O_7$, which reveals the most intense diffraction rings.}\label{fig:sem}
\end{figure}

Room temperature x-ray diffraction patterns of nanocrystalline and polycrystalline samples are shown in figure~\ref{fig:xrd}(a) and (b) respectively. Further structural studies have been carried out using Rietveld refinement of XRD data using FULLPROF suite. Rietveld analysis shows small fraction of impurity phase of $Y_2O_3$ in the polycrystalline sample. Other groups also noted significant amount of impurity phases such as $Y_2O_3$, $IrO_2$ and Ir in polycrystalline $Y_2Ir_2O_7$ prepared using solid state reaction~\cite{Shapiro,Disseler,Zhu}. Nevertheless, $Y_2O_3$ is diamagnetic with a small negative susceptibility and hence should have minimal contribution to the magnetic properties. The Rietveld refinement confirms the cubic pyrochlore structure with space group symmetry Fd-3m. The fitting parameters are a = 10.18{$\AA$}, V = 1055.8 ${\AA}^3$  (${\chi}^2$ = 1.61) and a = 10.17 {$\AA$}, V = 1053 ${\AA}^3$ (${\chi}^2$ = 3.25) for nanocrystalline and polycrstalline $Y_2Ir_2O_7$, respectively. The average crystallite size as calculated from the Debye-Scherer formula turns out to be 60 nm for nanocrystalline sample which is further confirmed by FE-SEM analysis. The FE-SEM micrograph (figure~\ref{fig:sem}(a), (b)) of nanocrystalline and bulk sample reveals uniform distribution of densely packed grains. The average particle size for bulk sample is 1 $\mu$m as shown in figure~\ref{fig:sem}(b). The Energy Dispersive X-ray spectroscopy (EDS) spectra, taken at a number of selected positions of the sample, shows the average expected presence of Y, Ir and O in nearly  stoichiometric ratio as shown in Table1.
\begin{table}[h!]
\centering
\caption{EDX results of nanocrystalline and polycrystalline $Y_2Ir_2O_7$ }
\begin{tabular}{|c|c|c|}
 \hline
 Elements & $Nanocrystalline (Atomic\%)$ & $Bulk (Atomic\%)$\\
 \hline
 $O $ & 73.9 & 64.9 \\
 \hline
 $Y $ & 13.0 & 18.7 \\
 \hline
 $Ir $ &13.0 & 16.4 \\
 \hline
\end{tabular}
\end{table}
The Transmission Electron Micrograph (TEM) in figure~\ref{fig:sem}(c) shows that the average particle size in nanocrystalline sample ranges between 50-60 nm which is in fair agreement with XRD and FE-SEM data. The Selected Area Electron Diffraction (SAED) pattern of the nanocrystalline sample is shown in figure~\ref{fig:sem}(d).

\begin{figure}[h]
\centering
\includegraphics[width=10cm]{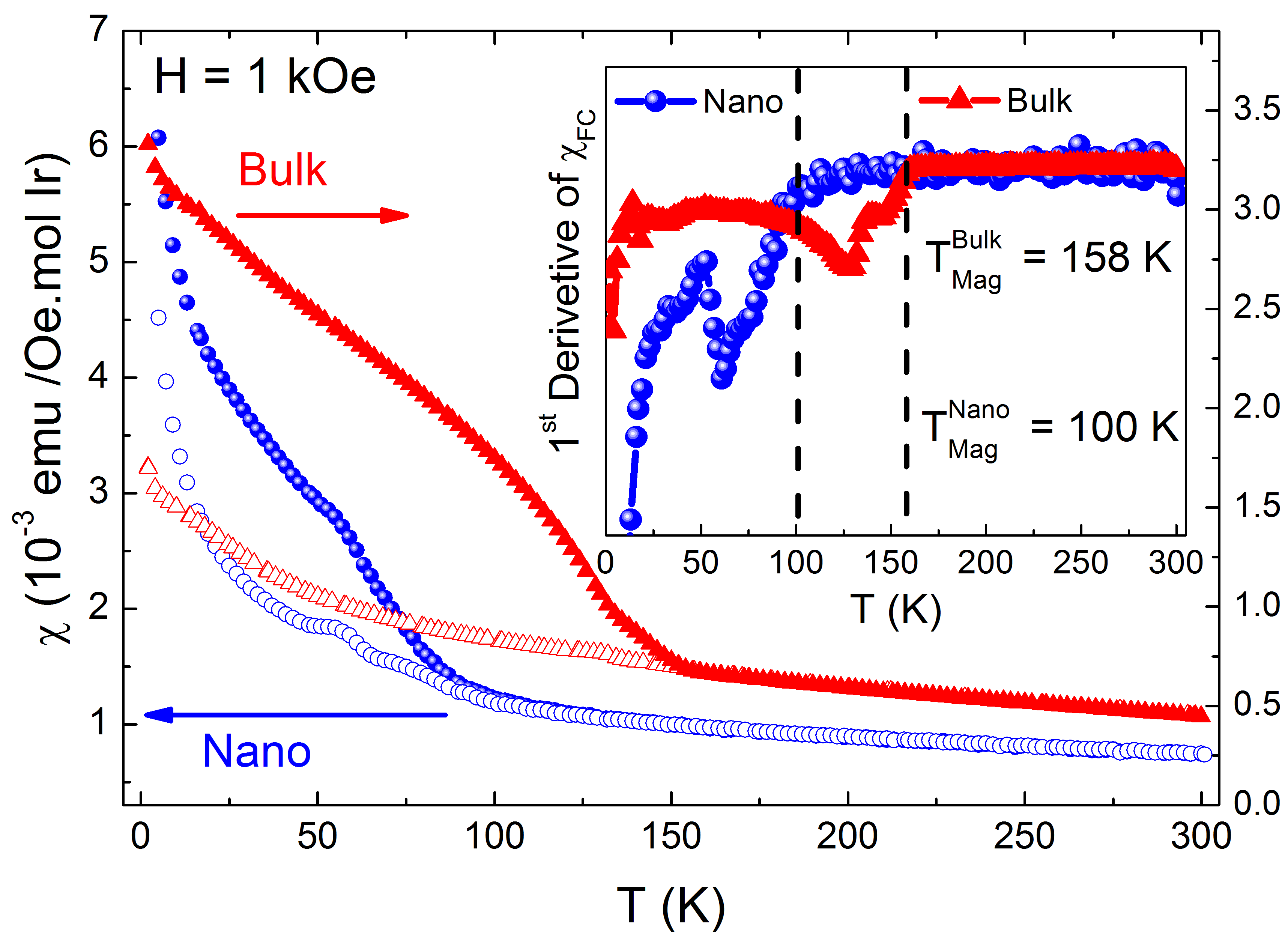}\\
\caption{Temperature dependence of ZFC and FC susceptibilities of nanocrystalline (blue open and solid circle) and polycrystalline (red open and solid triangle) $Y_2Ir_2O_7$ samples measured in a magnetic field H = 1 kOe. Inset shows first derivative of FC susceptibilities.}\label{fig:mag}
\end{figure}

Figure~\ref{fig:mag} shows the temperature dependent magnetic susceptibilities of nanocrystalline and polycrystalline $Y_2Ir_2O_7$ samples following zero field cooled (ZFC) and field cooled (FC) protocol at H = 1kOe. A bifurcation can be seen between ${\chi_{ZFC}}$ and ${\chi_{FC}}$ for polycrystalline and nanocrystalline samples near temperatures 160 K and 100 K respectively. To find out the magnetic transition temperature $T_{mag}$, we plotted the derivative of ${\chi_{FC}}$ (d${\chi_{FC}}$/dT) of both sample (inset of figure~\ref{fig:mag}) which show minima near 158 K for the bulk and 100 K for the nanocrystalline sample. The magnetic susceptibility data for the polycrystalline sample is consistent with previous studies~\cite{Disseler,Shapiro,Zhu,Taira,Fukazawa,Aito,Singh}. In nanocrystalline sample, below 50 K, the susceptibility increases very sharply. Interestingly, XRD pattern is unable to detect any impurity phase in nanocrystalline sample prepared by chemical route. The upturn in ${\chi_{FC}}$ and ${\chi_{ZFC}}$ could be suggestive of long range ferromagnetic ordering.
The magnetic susceptibility above $T_{Mag}$ obeys the Curie-Weiss law, $\chi = \frac{C}{T-\theta_{CW}} + \chi_0$, Where C, $\theta_{CW}$ and $\chi_{0}$ are the Curie constant, Weiss temperature and a temperature independent component of susceptibility, respectively. The parameters corresponding to the best fit turn out to be C = 0.19 emu K/Oe mol Ir, $\theta_{CW}$ = -128 K for polycrystalline and C = 0.41 emu K/Oe mol Ir, $\theta_{CW}$ = -252 K for nanocrystalline sample, the negative value of $\theta_{CW}$ suggesting an antiferromagnetic (AFM) correlation for both samples~\cite{Wan,Wan1}. The calculated effective magnetic moments are $\mu_{eff}$ = 0.61 $\mu_{B}$/f.u. for polycrystalline (close to literature value $\mu_{eff}$ = 0.54 $\mu_{B}$/f.u. ~\cite{Shapiro}) and $\mu_{eff}$ = 1.33 $\mu_{B}$/f.u. for nanocrystalline samples, respectively. These observed $\mu_{eff}$ are lower than the Hund's rule value $\mu_{eff}$ = 1.73 $\mu_{B}$/f.u. for S = 1/2. The enhancement in effective magnetic moment in nanocrystalline sample compared with polycrystalline sample can be attributed to the increased surface to volume ratio in nanocrystalline sample and the consequent enhanced contribution of uncompensated surface spins. It is possible that the strain induced by distortion at grain boundaries could change the Ir \textemdash O \textemdash Ir bond angle and facilitate ferromagnetic double exchange. We shall come back to this point shortly hereafter.
\begin{figure}[h]
\centering
\includegraphics[width=10cm]{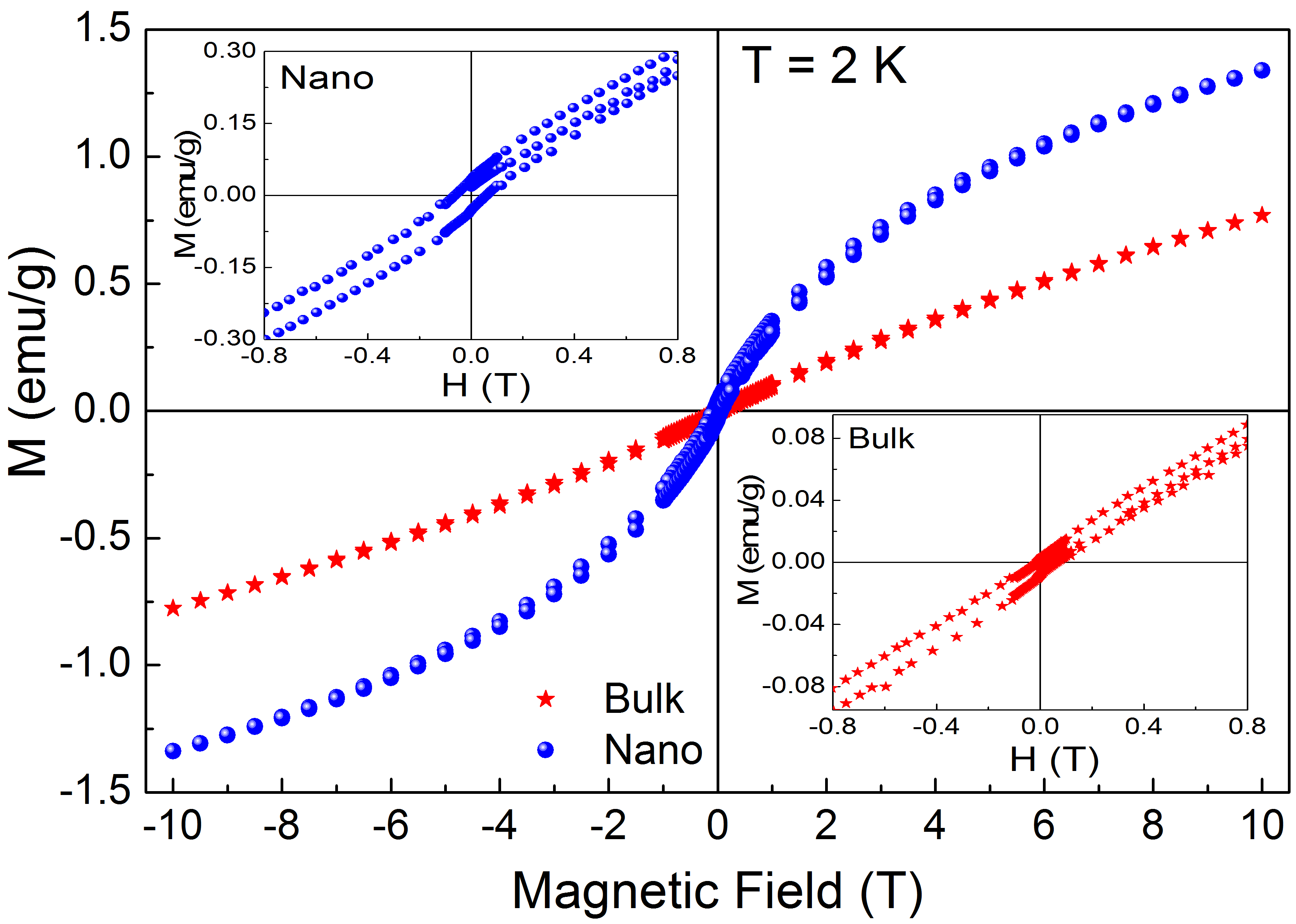}\\
\caption{Magnetization as a function of field at 2 K of nanocrystalline (blue solid circle) and bulk (red solid star) $Y_2Ir_2O_7$ samples. Upper left inset and lower right inset shows low field magnetization data for nano and bulk, respectively.}\label{fig:hyst}
\end{figure}

To further explore the magnetic properties of the polycrystalline and nanocrystalline $Y_2Ir_2O_7$ samples, magnetic hysteresis loops have been measured at 2 K as shown in figure~\ref{fig:hyst}. For both samples, weak ferromagnetic (FM) like hysteretic behaviour is observed with coercive field $H_C$ being 375 Oe and 610 Oe, for polycrystalline and nanocrystalline sample, respectively. There is no sign of magnetic saturation up to 10 T. The small hysteresis in M vs H data may originate from short range type FM ordering at low temperatures while the sharp enhancement of magnetization at low temperature might be due to canting of the moments from the AFM all-in/all-out state. The overall magnetic structure of the nanocrystalline $Y_2Ir_2O_7$ can be considered as a core/shell system, where the inner part of the particle is in AFM phase and the surface is a disordered ferromagnet.

\begin{figure}[h]
\centering
\includegraphics[width=10cm]{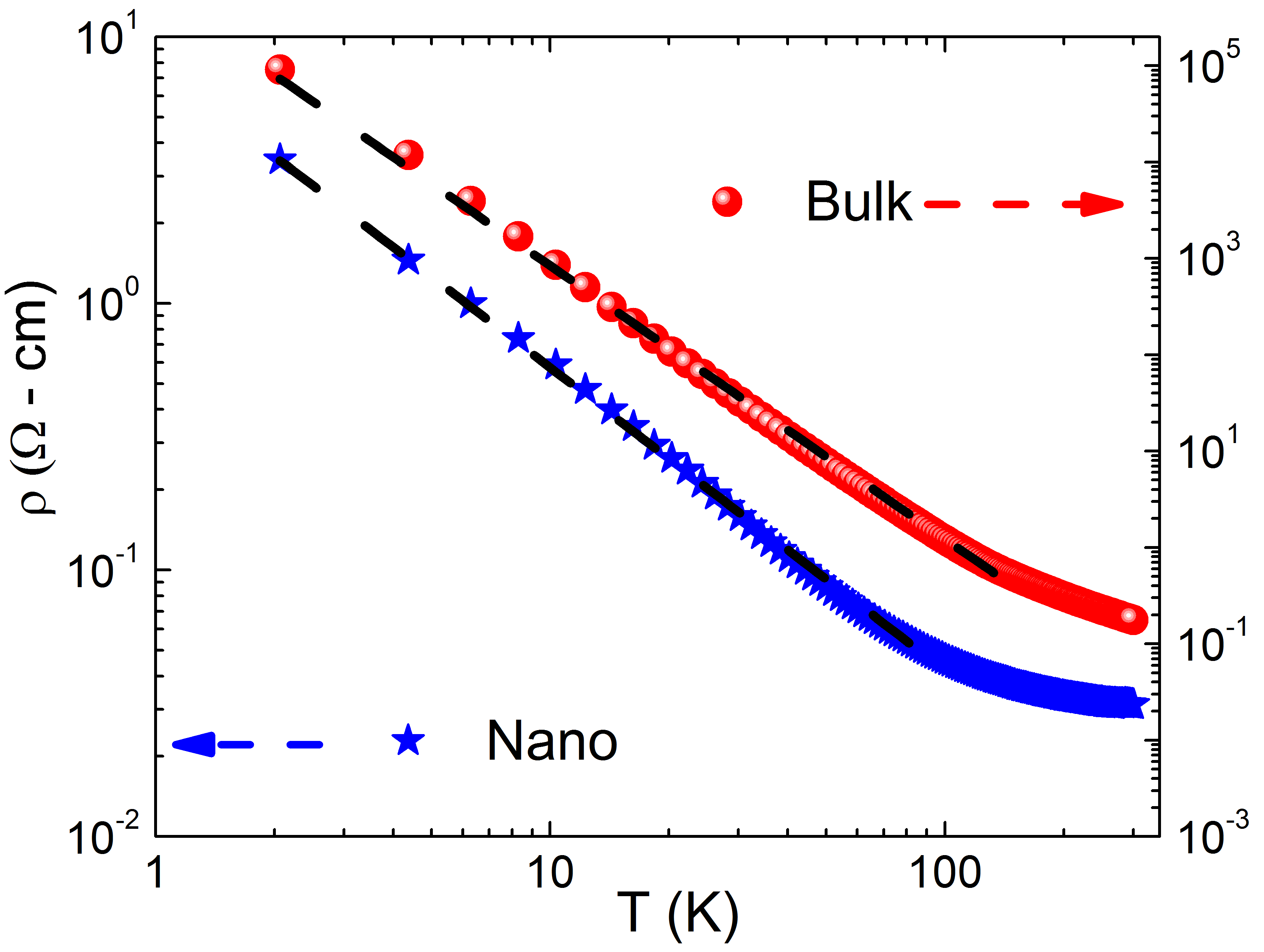}\\
\caption{Temperature dependence of zero field electrical resistivity $\rho (T)$ measured during heating for nanocrystalline (blue solid star) and bulk (red solid circle) $Y_2Ir_2O_7$ samples showing power law behaviour of both samples up to near their magnetic transition temperature.}\label{fig:res}
\end{figure}

The temperature dependent resistivity (figure~\ref{fig:res}) shows insulating behaviour consistent with previous study for polycrystalline sample~\cite{Yanagishima,Fukazawa,Soda,Aito,Singh,Disseler,Zhu}. Strikingly, there is at least 4 orders of magnitude enhancement of electrical conductivity at low temperature in the nanocrystalline sample compared to the bulk with the resistivity values at 2 K being 9.1$\times10^{4} \Omega$-cm and 3.45 $\Omega$-cm for bulk and nanocrystalline sample, respectively. It turns out that the temperature dependence of zero field resistivity follows power law behaviour $\rho = \rho_0T^{-n}$ below $T_{mag}$ (figure~\ref{fig:res}) with exponents n = 2.8 and 1.1 for polycrystalline and nanocrystalline samples, respectively. Similar power law driven electronic transport has been observed for polycrystalline $ Y_2Ir_2O_7 $ by other groups\cite{Disseler}. As discussed earlier, there is possibility that reduction in particle size increases the valence state of Ir from $Ir^{4+}$ to $Ir^{+5}$ facilitating double exchange. For stoichiometric A-227 compounds, the $Ir^{4+}$ has an unpaired $J_{eff}$ = 1/2 electron that is localized due to the electron-electron interaction~\cite{Kim,Pesin,Wan}. On the other hand, $Ir^{+5}$ has an empty $J_{eff}$ = 1/2 level which could allow hopping of the $J_{eff}$ = 1/2 electron from the nearby $Ir^{+4}$, leading to the delocalization of electrons and enhancement of electrical conductivity.
\begin{figure}[h]
\centering
\includegraphics[width=10cm]{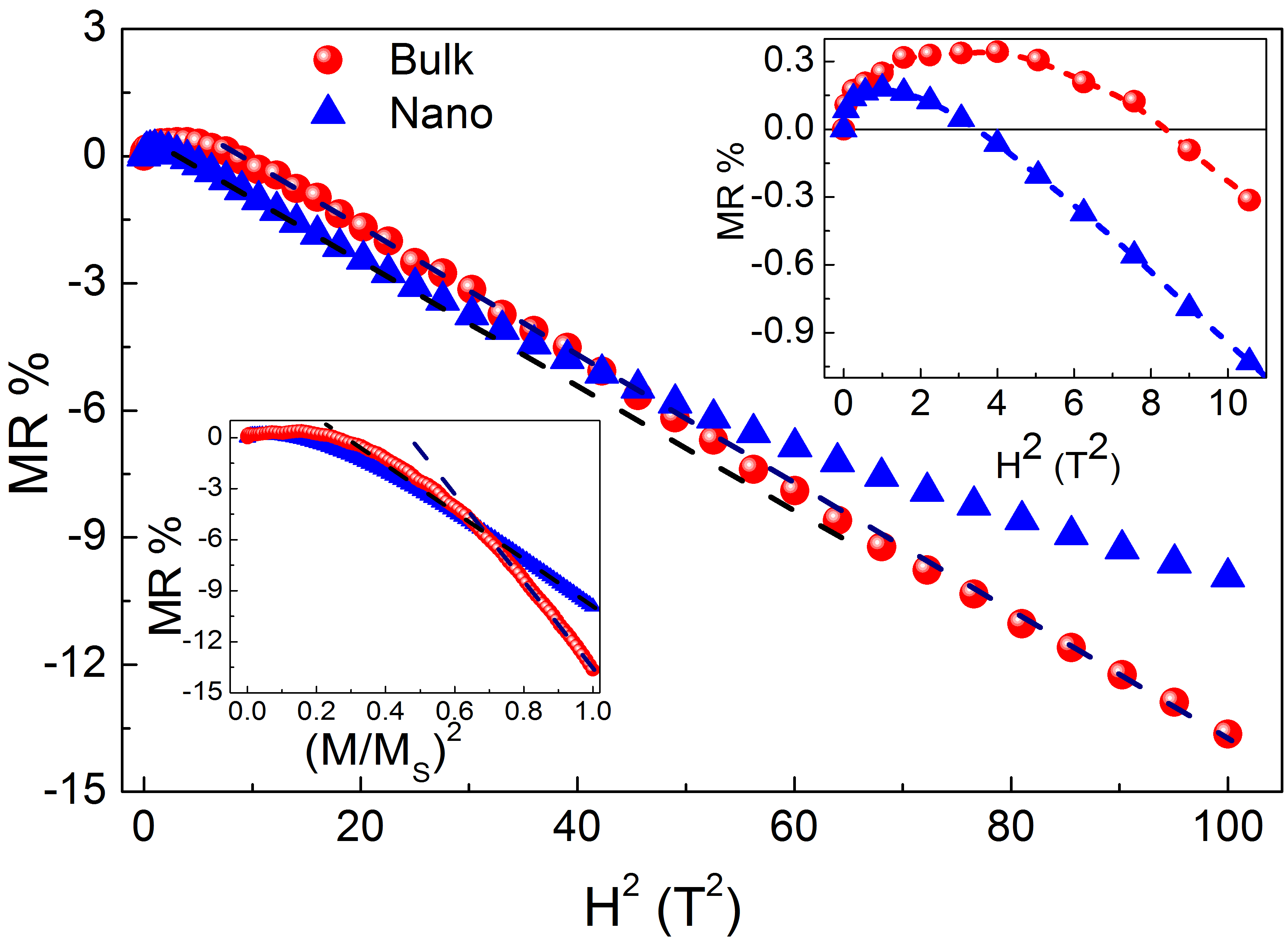}\\
\caption{Magnetoresistance (MR) as a function of applied magnetic field for polycrystalline (red solid circle) and nanocrystalline (blue solid triangl) $Y_2Ir_2O_7$, recorded at temperature 2K, shows quadratic field dependence of MR for bulk while at high magnetic field there is a clear deviation for nanocrystalline sample. Upper inset is the expanded view of the low field portion of the MR data, while lower inset shows linear variation of MR with square of reduced magnetization.}\label{fig:mr}
\end{figure}

Further signature of double exchange mechanism is found in the electronic transport behavior in the presence of magnetic field at low temperature. The magnetoresistance is positive at low field for both samples (Top Inset, figure~\ref{fig:mr}) which becomes negative at high magnetic field. The positive MR at low field could be attributed to weak antilocalization which is expected for systems with strong spin orbit coupling although, surprisingly, so far unreported. The reason could be significant amount of impurity phases in the polycrystalline samples prepared by other groups. The peak value of positive MR in the bulk is considerably higher than that in the nanocrystalline sample implying reduction of spin-orbit coupling in the latter. The observed value of negative MR is 14$\%$ and 10$\%$ for bulk and nanocrystalline $Y_2Ir_2O_7$, respectively at maximum field 10 T. There is no sign of saturation in MR up to maximum applied magnetic field 10 T. Figure~\ref{fig:mr} shows quadratic field dependence of MR for bulk sample while for nanocrystalline sample, MR deviates from quadratic field dependence at high magnetic field. Interestingly the MR for both samples scales with magnetization which is a characteristic of double exchange systems near magnetic transition (Bottom Inset, figure~\ref{fig:mr}).

\begin{figure}[h]
\centering
\includegraphics[width=10cm]{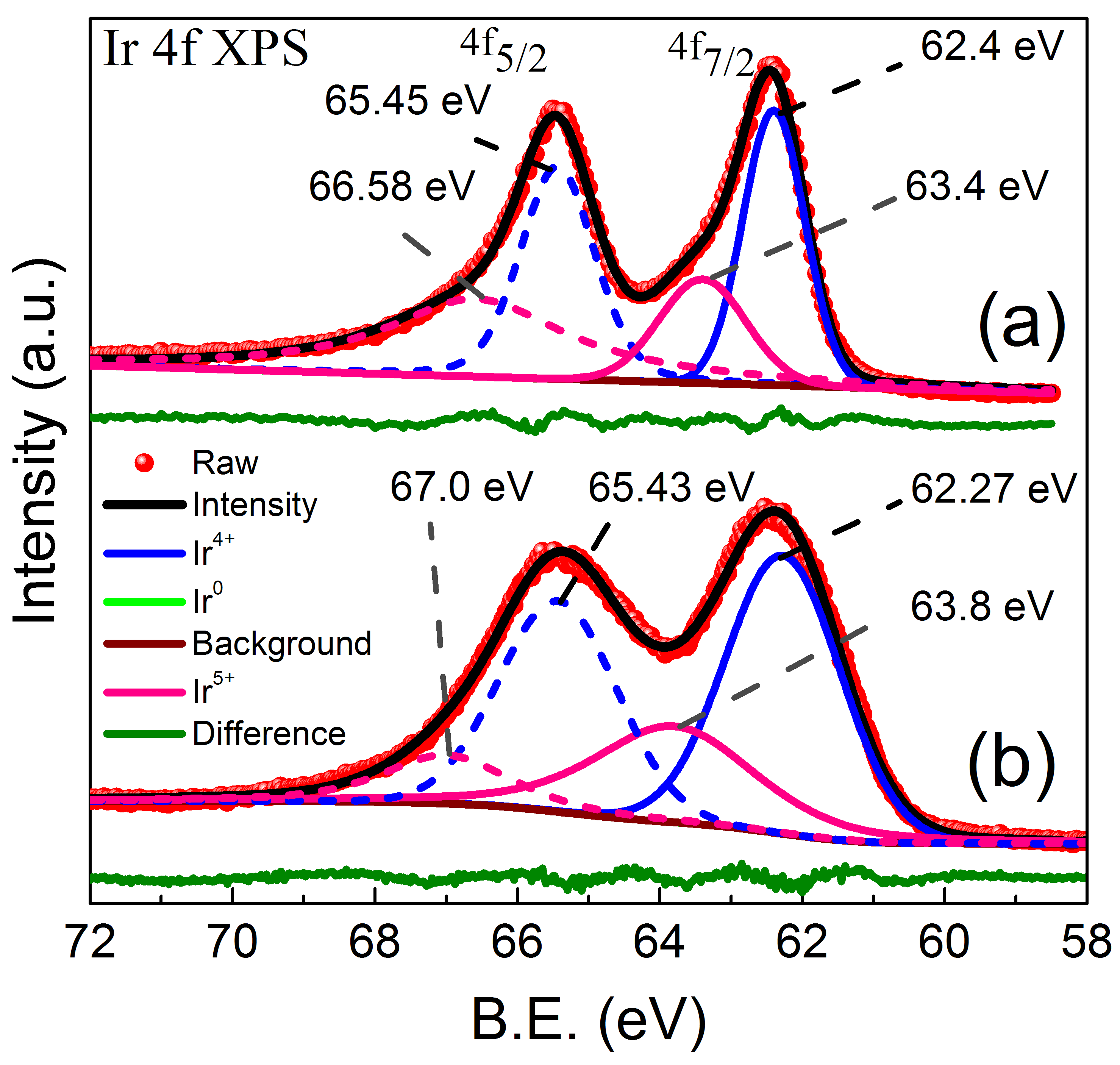}\\
\centering
\includegraphics[width=10cm]{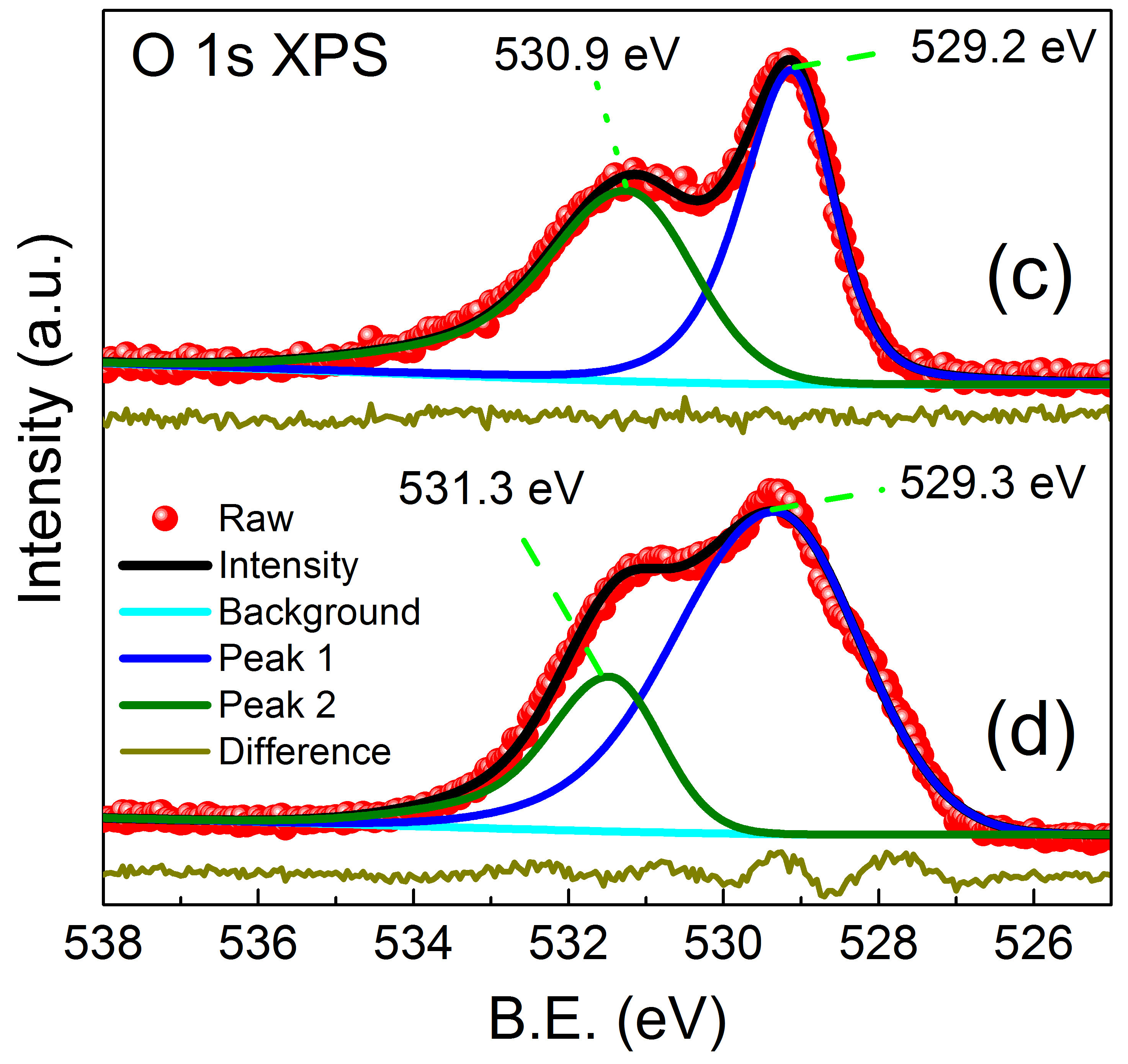}\\
\caption{XPS spectra of Ir 4f and O 1s line of polycrystalline(a),(c) and nanocrystalline (b)(d) $Y_2Ir_2O_7$, respectively.}\label{fig:xps}
\end{figure}

So far we have underlined that the possible reason behind the emergence of enhanced ferromagnetism and electrical conductivity in nanocrystalline sample is mixed oxidation state of Ir. In oreder to verify this proposition we characterize the oxidation state of Ir using X-ray Photoelectron Spectroscopy (XPS). XPS spectra for polycrystalline and nanocrystalline samples are shown in figure~\ref{fig:xps}. We have fitted our XPS data using three sets of iridium components by asymetric Guass-Lorentz sum function with shirley background using XPS peakfit4.1 software. The experimental data indicates that the two valence states of Ir are needed for a proper fit of XPS data in case of nanocrystalline as well as polycrystalline sample. The peaks identified as $4f_{7/2}$ at around 62 eV (blue solid line) and $4f_{5/2}$ at 65 eV (blue dashed line) in figure~\ref{fig:xps}(a) and ~\ref{fig:xps}(b) respectively, are attributed to the $4^{+}$ valence state of Ir, which is similar to the $IrO_2$ single crystal~\cite{Wertheim}. On the other hand, the two distinct additional features at around 63.5 eV (pink solid line) and 67 eV (pink dashed line) represents the contribution of $Ir^{5+}$ oxidation state for both the compounds. The presence of  $Ir^{5+}$ in the polycrystalline sample might be due to nonstoichiometry (i.e. deficiency of metal elements such as Y/Ir or excess of oxygen), which is consistent with previous reports~\cite{Zhu}. Interestingly, for nanocrystalline sample, the peak area for $Ir^{5+}$ is found to increase by 37\% as compared to bulk suggesting larger fraction of $Ir^{5+}$ in the former.

 Further support for the mixed valence of Ir is found in the O 1s spectra ( which usually exhibits single peak centered at 529.3 eV~\cite{Wertheim}) shown in figure~\ref{fig:xps}(c) and ~\ref{fig:xps}(d). We observe a doublet having energies 529 eV and 531 eV for both compounds. Multiple oxidation states affect the local environment of Ir\textemdash O bonds leading to the doublet feature. We also performed XPS measurement on $Y 3d_{3/2}$ region (not shown here). The $3d_{3/2}$ peak shows a single feature at 156.42 eV and 158.34 eV, which is close to value of $Y_2O_3$, suggesting that only $Y^{3+}$ is present in the material.

\begin{figure}[h]
\centering
\includegraphics[width=10cm]{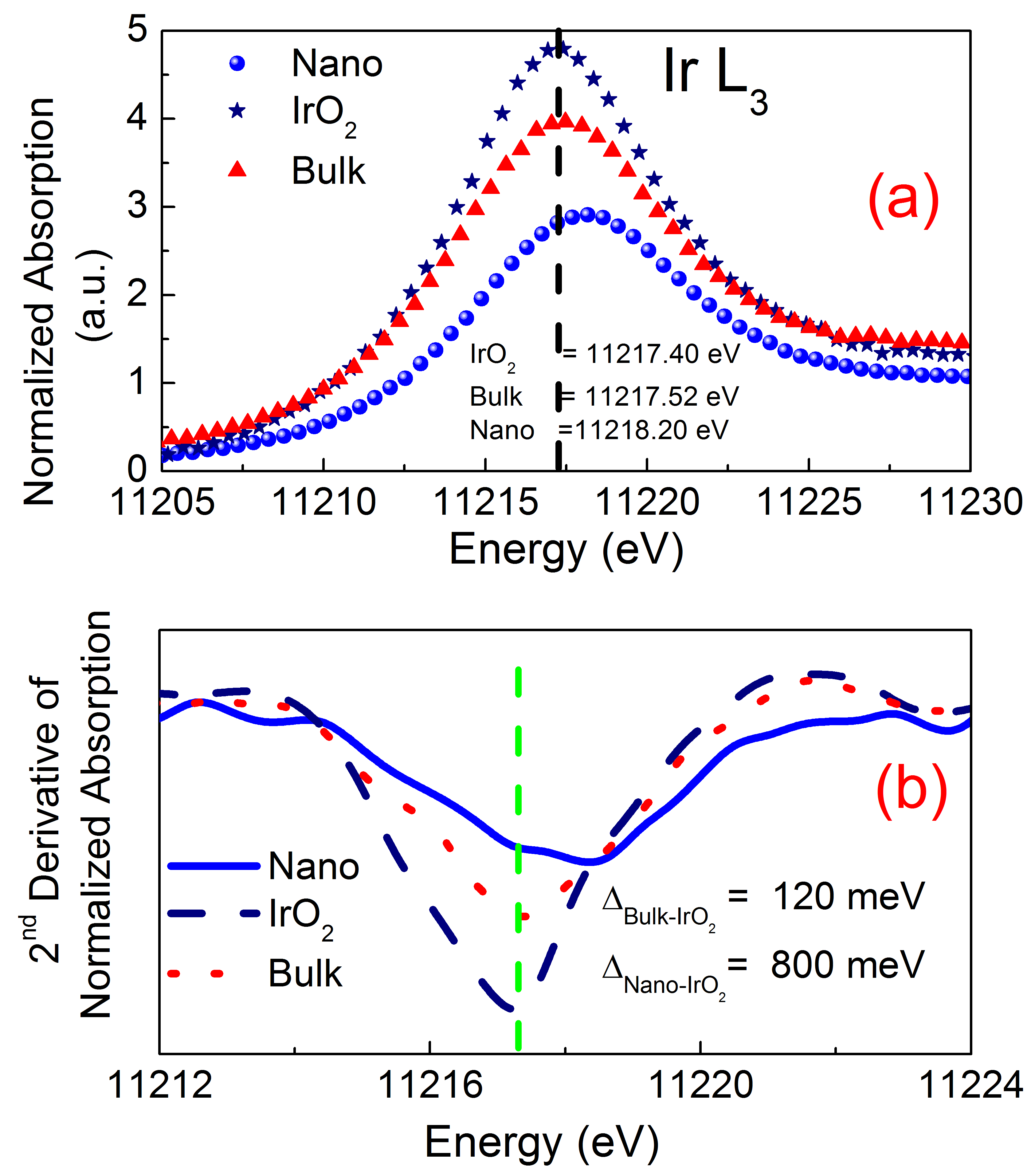}\\
\caption{XANES spectra of Ir $L_{III}$ white line (a)normalized  and (b) second derivative, of polycrystalline, nanocrystalline and reference sample $IrO_2$}\label{fig:xanes}
\end{figure}

As a further investigation into determining the change in valence state of Ir, X-ray Absorption Near Edge Structure (XANES) spectroscopy is also carried out. Figure~\ref{fig:xanes}(a) shows $Ir L_{III}$-edge XANES spectra along with reference material $IrO_2$. All the samples show very intense white lines suggesting a large local density of 5d states. However, clear differences can be observed in the XANES profiles indicating differences in the electronic structure among these systems. There is some broadening of the white line with increasing energy. The position of the white-line feature  for polycrystalline $Y_2Ir_2O_7$ ($Ir^{4+}$) and nanocrystalline $Y_2Ir_2O_7$ ($Ir^{5+}$) sample shifts towards higher energies when compared to $IrO_2$ ($Ir^{4+}$) reference sample. For nanocrystalline sample the shift is close to 1 eV, much higher compared to the bulk, strongly suggesting existence of higher valence state i.e. +5 ~\cite{Clancy}. In addition, the spectral shape of samples with nominal $Ir^{4+}$ and $Ir^{5+}$ states shows an unresolved double-peak feature including a shoulder on the low-energy side of the peak. This structure can be more clearly visible in the profile of the second derivative of the spectra for the nanocrystalline sample, as shown in the figure~\ref{fig:xanes}(b). 

\begin{figure}[h]
\centering
\includegraphics[width=10cm]{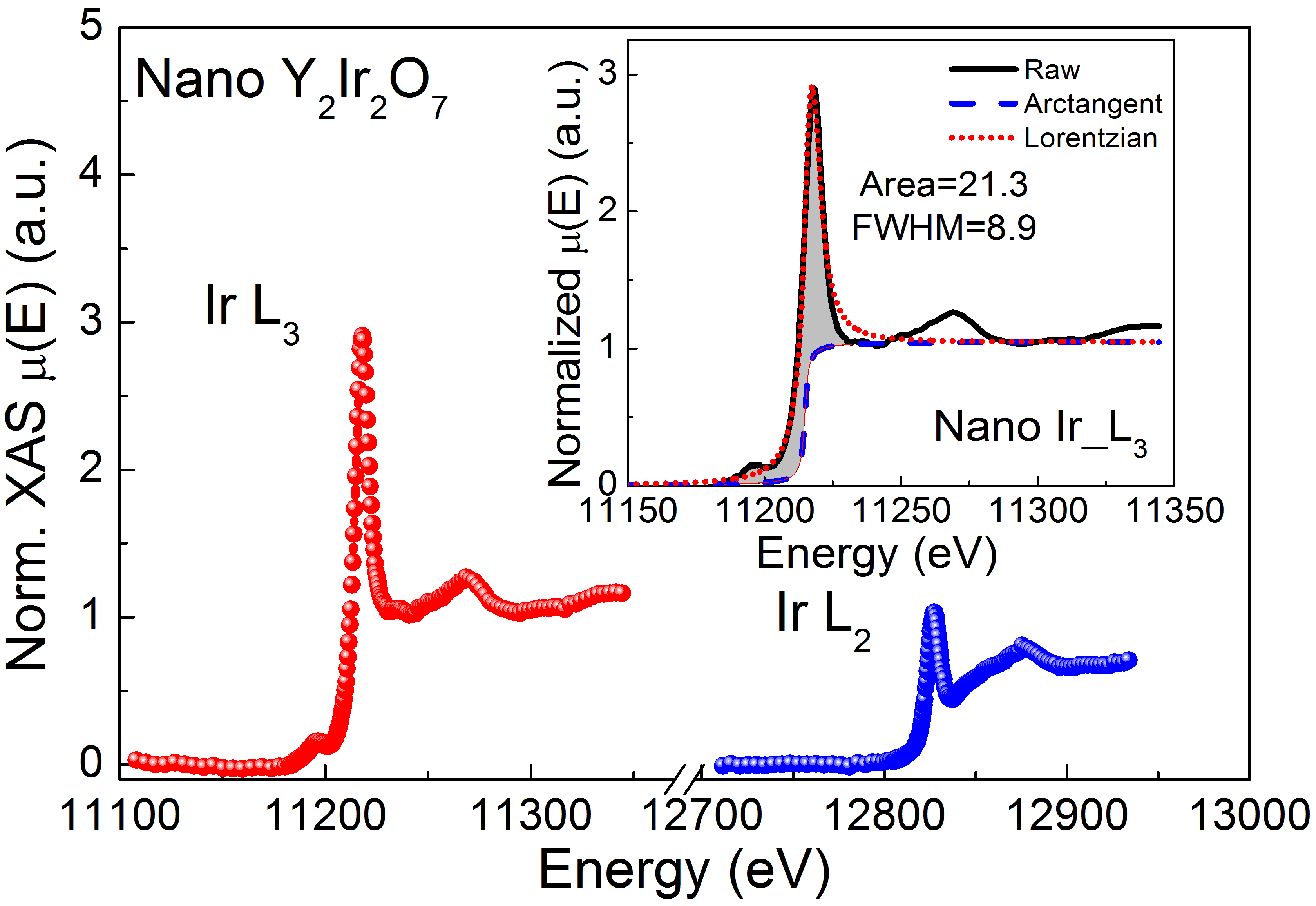}\\
\caption{XANES spectra of nanocrystalline sample collected at Ir $L_{III}$ and $L_{II}$ white line absorption edges. Inset: Experimental determination of the white line intensity at the Ir $L_{III}$ absorption edge in nanocrystalline $Y_2Ir_2O_7$. The black solid line represents the experimental data, while the blue dashed line represents the arctangent function. The red short dot line represents the best fit to the data using a Lorentzian + arctangent fit function.}\label{fig:xanes1}
\end{figure}

G. van der Laan and Thole, proposed that the intensity ratio of the white line features observed in XAS measurements at the $L_{III}$ and $L_{II}$ absorption edges is directly proportional to the expectation value of the spin-orbit operator $<L.S>$  = $<\sum_i l_i.s_i>$ ~\cite{Laan}. The ratio of the integrated white line intensity recorded at edges $L_{III}$ and $L_{II}$ is known as branching ratio $BR = \frac{I_{L3}}{I_{L2}} = \frac{2+r}{2-r}$, where $r = \frac{<L.S>}{<n_h>}$. Here $<n_h>$ = 5 is the number of holes in related compounds. We calculated spin orbit interaction from normalized XANES spectra recorded at $L_{III}$ and $L_{II}$ white line positions (figure~\ref{fig:xanes1}). We follow the method employed by Clancy et al. ~\cite{Clancy} to obtain the white line intensities as shown in the inset of figure~\ref{fig:xanes1} for Ir$L_{III}$ edge of nanocrystalline $Y_2Ir_2O_7$. The experimental data is represented by the black solid line while the blue dashed line represents the continuum edge-step described by an arctangent function located at the center of the absorption edge with unit height. This function is subtracted from the raw data, leaving only the white line contribution (red short dot line) in approximately Lorentzian shape.

\begin{table}[h!]
\centering
\caption{Summary of results obtained from XANES measurements}
\begin{tabular}{|c|c|c|c|c|}
 \hline
 compound & $I_{L3}(numeric)$ & $I_{L2}(numeric)$& BR& $<L.S>[\hbar^{2}]$\\
 \hline
 $IrO_2$ & 33.6 & 5.1 & 6.6 & 3 \\
 \hline
 Bulk $Y_2Ir_2O_7$ & 29.0 & 5.2 & 5.6 & 2.7 \\
 \hline
 Nano $Y_2Ir_2O_7$ & 21.3 & 5.0 & 4.3 & 2.2 \\
 \hline
\end{tabular}
\end{table}

We found that the branching ratio for $IrO_2$, polycrystalline and nanocrystalline $Y_2Ir_2O_7$ are 6.6, 5.6 and 4.3 respectively. These branching ratios for all compounds are larger than the statistical branching ratio of 2, which is a characteristic of Ir based metallic systems~\cite{Jeon,Wilhelm}. This result indicates a strong relativistic spin orbit coupling effect for all the samples albeit with significant relative reduction in strength for nanocrystalline sample compared to the bulk (see the above Table 2).

\section{CONCLUSIONS}

Now we summarize the principal aspects of the experimental results and the conclusions drawn. (1) There is an enhancement of effective magnetic moment in the nanocrystalline sample compared to the bulk. The XPS result suggests greater prevalence of the coexistence of $Ir^{5+}$ and $Ir^{4+}$ in nanocrystalline sample than in polycrystalline sample. The resulting mixed valence state could favor a double exchange mechanism leading to the enhancement of the effective ferromagnetic contribution. The weakening of spin orbit interaction in nanocrystalline sample as revealed by the XANES result could also increase the effective magnetic moment. (2) We also observe a striking orders of magnitude enhancement in electrical conductivity in the nanocrystalline sample compared to the bulk. The weakening of spin orbit coupling strength would lead to an increase in the kinetic energy term leading to enhanced conductivity. Both enhanced double exchange and lower spin orbit coupling should favor higher electrical conductivity. We emphasize that reducing the particle size should play similar role as applying chemical pressure and thus opens up a new direction for research on Pyrochlore Iridates.

 \section*{ACKNOWLEDGMENTS}
 The authors would like to acknowledge RRCAT, Indore, India for providing the XANES measurement facility. VKD is thankful to Dr. S. N. Jha and C. Nayak of RRCAT, Indore for valuable discussions.


\section*{References}
\begin{thebibliography}{99}
\bibitem{Wan} X. Wan, Ari M. Turner, Ashvin Vishwanath, and Sergey Y. Savrasov 2011 \textit{Phys. Rev. B} {\bf{83 205101}}.

\bibitem{Kim} B. J. Kim, Hosub Jin, S. J. Moon, J.-Y. Kim, B.-G. Park, C. S. Leem, Jaejun Yu, T. W. Noh, C. Kim, S.-J. Oh, J.-H. Park, V. Durairaj, G. Cao, and E. Rotenberg 2008 \textit{Phys. Rev. Lett.} {\bf{101 076402}}.

\bibitem{Pesin} Dmytro Pesin and Leon Balents 2010 \textit{Nature Physics} {\bf{6 376-381}}.

\bibitem{Clancy} J. P. Clancy, N. Chen, C. Y. Kim, W. F. Chen, K. W. Plumb, B. C. Jeon, T. W. Noh, and Young-June Kim 2012 \textit{Phys. Rev. B} {\bf{86 195131}}.

\bibitem{Yang}B.-J. Yang and Y. B. Kim 2010 \textit{Phys. Rev. B} {\bf{82 085111}}.

\bibitem{Bhattacharjee}Subhro Bhattacharjee, Sung-Sik Lee and Yong Baek Kim 2012 \textit{New Journal of Physics} {\bf{14 073015}}.

\bibitem{William}William Witczak-Krempa, Gang Chen, Yong Baek Kim, and Leon Balents 2014 \textit{Annu. Rev. Condens. Matter Phys.} {\bf{5 57}}.

\bibitem{Kargarian}M. Kargarian, J.Wen, and G. A. Fiete 2011 \textit{Phys. Rev. B} {\bf{83 165112}}.

\bibitem{Machida}Y. Machida, S. Nakatsuji, S. Onoda, T. Tayama, and T. Sakakibara 2010 \textit{Nature} {\bf{463 210}}.

\bibitem{Go}A. Go, W.Witczak-Krempa, G. S. Jeon, K. Park, and Y. B. Kim 2012 \textit{Phys. Rev. Lett.} {\bf{109 066401}}.

\bibitem{Chen}G. Chen and M. Hermele 2012 \textit{Phys. Rev. B} {\bf{86 235129}}.

\bibitem{William1}William Witczak-Krempa and Yong Baek Kim 2012 \textit{Phys. Rev. B} {\bf{85 045124}}.

\bibitem{Shinaoka}H. Shinaoka, T. Miyake, and S. Ishibashi 2012 \textit{Phys. Rev. Lett.} {\bf{108 247204}}.

\bibitem{Bogdanov}N. A. Bogdanov, R. Maurice, I. Rousochatzakis, J. van den Brink, and L. Hozoi 2013 \textit{Phys. Rev. Lett.} {\bf{110 127206}}.

\bibitem{Taira} N. Taira, M. Wakeshima, and Y. Hinatsu 2001 \textit{J. Phys.: Condens. Matter} {\bf{13 5527}}.

\bibitem{Disseler2} S. M. Disseler, Chetan Dhital, T. C. Hogan, A. Amato, S. R. Giblin, Clarina de la Cruz, A. Daoud-Aladine, Stephen D. Wilson, and M. J. Graf 2012 \textit{Phys. Rev. B} {\bf{85 174441}}.


\bibitem{Guo}H. Guo, K. Matsuhira, I. Kawasaki, M.Wakeshima, Y. Hinatsu, I. Watanabe, and Z.-a. Xu 2013 \textit{Phys. Rev. B} {\bf{88 060411(R)}}.

\bibitem{Tomiyasu}K. Tomiyasu, K. Matsuhira, K. Iwasa, M. Watahiki, S. Takagi, M. Wakeshima, Y. Hinatsu, M. Yokoyama, K. Ohoyama, and K. Yamada 2012 \textit{J. Phys. Soc. Jpn.} {\bf{81 034709}}.

\bibitem{MacLaughlin}D. E. MacLaughlin, Y. Ohta, Y. Machida, S. Nakatsuji, G. M. Luke,K. Ishida, R. H. Heffner, L. Shu, andO.O. Bernal 2009 \textit{Physica B} {\bf{404 667}}.

\bibitem{Disseler} S. M. Disseler,Chetan Dhital, A. Amato, S. R. Giblin, Clarina de la Cruz, Stephen D. Wilson, and M. J. Graf 2012  \textit{Phys. Rev. B} {\bf{86 014428}}.

\bibitem{Shapiro} M. C. Shapiro, Scott C. Riggs, M. B. Stone, C. R. de la Cruz, S. Chi, A. A. Podlesnyak, and I. R. Fisher 2012 \textit{Phys. Rev. B} {\bf{85 214434}}.

\bibitem{Disseler1} Steven M. Disseler 2014 \textit{Phys. Rev. B} {\bf{89 140413(R)}}.
\bibitem{Zhu} W. K. Zhu, M. Wang, B. Seradjeh, Fengyuan Yang, and S. X. Zhang 2014  \textit{Phys. Rev. B} {\bf{90 054419}}.

\bibitem{Singh} R. S. Singh, V. R. R. Medicherla, Kalobaran Maiti, and E. V. Sampathkumaran 2008 \textit{Phys. Rev. B} {\bf{77 201102(R)}}.

\bibitem{Fukazawa} H. Fukazawa and Y. Maeno 2002 \textit{J. Phys. Soc. Jpn.} {\bf{71  2578 }}.

\bibitem{Aito} N. Aito, M. Soda, Y. Kobayashi and M. Sato 2003 \textit{J. Phys. Soc. Jpn.} {\bf{72 1226}}.

\bibitem{Soda} M. Soda, N. Aito, Y. Kurahashi, Y. Kobayashi, M. Sato 2003  \textit{Physica B} {\bf{329 1071–1073}}.

\bibitem{Yanagishima} Daiki Yanagishima and Yoshiteru Maeno 2001  \textit{J. Phys. Soc. Jpn. } {\bf{70 2880-2883}}.

\bibitem{http}http://www.rrcat.gov.in/technology/accel/srul/beamlines/exafsscan.html.
\bibitem{Wan1} X. Wan, Ashvin Vishwanath, and Sergey Y. Savrasov 2012  \textit{Phys. Rev. Lett.} {\bf{108 146601}}.

\bibitem{Wertheim} G. K. Wertheim and H. J. Guggenheim 1980 \textit{Phys. Rev. B} {\bf{22 4680}}.

\bibitem{Laan} G. van der Laan and B. T. Thole 1988 \textit{Phys. Rev. Lett.} {\bf{60 1977}}.


\bibitem{Jeon}Y. Jeon, Boyun Qi, F. Lu, and M. Croft 1989 \textit{Phys. Rev. B} {\bf{40 1538}}.


\bibitem{Wilhelm}F. Wilhelm, P. Poulopoulos, H. Wende, A. Scherz, K. Baberschke, M. Angelakeris, N. K. Flevaris, and A. Rogalev 2001 \textit{Phys. Rev. Lett.} {\bf{87  207202}}.



\end{thebibliography}
\end{document}